# Broadband NIR photon upconversion generates NIR persistent luminescence for bioimaging

*Shuting Yang,[1,‡] Bing Qi,[1,‡] Mingzi Sun,[2] Wenjing Dai,[1] Ziyun Miao,[1] Wei Zheng,[3] Bolong Huang,[2,*] Jie Wang[1,*]*

[1] The Key Lab of Health Chemistry & Molecular Diagnosis of Suzhou, College of Chemistry, Chemical Engineering & Materials Science, Soochow University, Suzhou 215123, China.

[2] Department of Applied Biology and Chemical Technology, The Hong Kong Polytechnic University, Hung Hom, Kowloon, Hong Kong SAR, China.

[3] CAS Key Laboratory of Design and Assembly of Functional Nanostructures, Fujian Key Laboratory of Nanomaterials, and State Key Laboratory of Structural Chemistry, Fujian Institute of Research on the Structure of Matter, Chinese Academy of Sciences, Fuzhou, Fujian 350002, China.

**ABSTRACT** Upconversion persistent luminescence (UCPL) phosphors that can be directly charged by near-infrared (NIR) light have gained considerable attention due to their promising applications ranging from photonics to biomedicine. However, current lanthanide-based UCPL phosphors show small absorption cross-sections and low upconversion charging efficiency. The development of UCPL phosphors faces challenges of lacking flexible upconversion charging pathways and poor design flexibility. Herein, we discovered a new lattice defect-mediated broadband photon upconversion process and the accompanied NIR-to-NIR UCPL in Cr-doped zinc gallate nanoparticles. The zinc gallate nanoparticles can be directly activated by broadband NIR light in the 700-1000 nm range to produce persistent luminescence at about 700 nm, which is also readily enhanced by rationally tailoring the lattice defects in the phosphors. This proposed UCPL phosphors achieved a signal-to-background ratio of over 200 in bioimaging by efficiently avoiding interference from autofluorescence and light scattering. Our findings reported the lattice defect-mediated photon upconversion for the first time, which significantly expanded the horizons for the flexible design of NIR-to-NIR UCPL phosphors toward broad applications.

Photon upconversion that converts multiple low-energy photons into high-energy ones has attracted enormous attention in fields including bioimaging,[1] phototherapy,[2] optogenetics,[3] and photocatalysis.[4-7] In the past decade, researchers have attempted to integrate the upconversion process with persistent luminescence based on different approaches.[8-10] Persistent luminescence is another optical phenomenon whereby luminescence remains after excitation ceases.[11-14] Upconversion persistent luminescence (UCPL) phosphors trap the NIR photon energy in lattice defects and gradually release the trapped energy radiatively after switching off excitation, possessing unprecedented advantages in many areas.[15] For instance, UCPL nanoprobes are capable of continuously releasing the trapped energy for autofluorescence-free bioimaging and sustained phototherapy after a single dosage of NIR light illumination.[16] Moreover, UCPL nanoprobes can be easily reactivated *in vivo* by NIR light after the exhaustion of trapped energy, enabling important applications of UCPL phosphors in imaging and treating deep-seated tumors.[10, 17] Therefore, developing UCPL is of great value, which involves disciplinary efforts and contributions from photonics, materials science, and engineering.

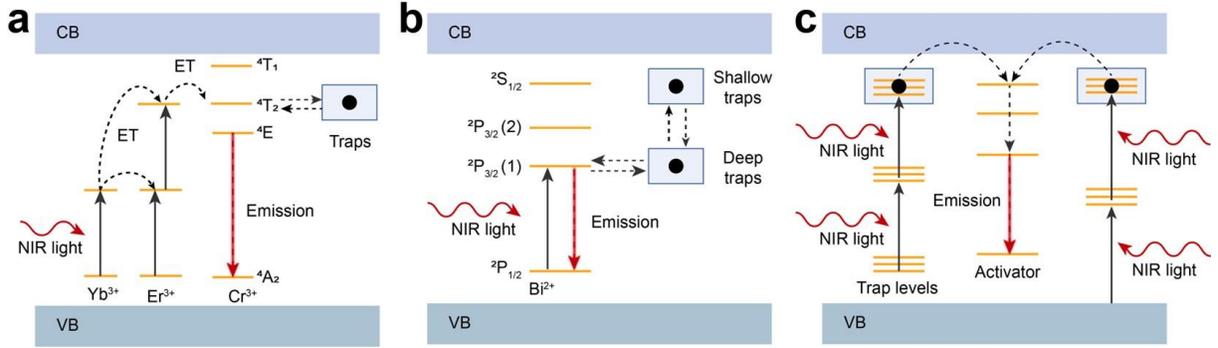

**Figure 1.** Schematic illustration of the reported and proposed UCPL process. (a) Lanthanides-mediated photon upconversion.[8] Under 980 nm excitation, the classic upconversion ion pair $Yb^{3+}$/$Er^{3+}$ absorbs two photons and passes the energy to $Cr^{3+}$, followed by the delivery of the excited electrons to electron traps by $Cr^{3+}$. After excitation ceases, the back-transfer of electrons from traps to $Cr^{4+}$ produces persistent luminescence. (b) Activators and electron traps co-mediated photon upconversion.[10] The doped $Bi^{2+}$ absorbs NIR photons and passes the excited electrons to deep traps, followed by the migration of electrons to shallow traps under thermal stimulation. After excitation ceases, the recombination of the trapped electrons with $Bi^{3+}$ generates persistent luminescence. (c) The proposed trap level-mediated photon upconversion. The electron in the occupied trap levels above VBM (left panel) or electron in VB (right panel) sequentially absorbs NIR photons and transitions to the empty trap levels beneath CBM, in which the continuously distributed trap level serves as the intermediary energy levels. After excitation ceases, the migration of the trapped electrons to activators produces persistent luminescence.

Pan's group pioneered UCPL in 2014 by doping the classic $Yb^{3+}$-$Er^{3+}$ pair into $Zn_3Ga_2GeO_8$:Cr persistent luminescence phosphor (Figure 1a).[8] The $Yb^{3+}$-$Er^{3+}$ pair absorbs two NIR photons (980 nm) and transfers the energy to $Cr^{3+}$, followed by the delivery of excited electrons from its delocalized state to electron traps by $Cr^{3+}$. After excitation ceases, the back-transfer of electrons from traps to the delocalized $Cr^{4+}$ produces persistent luminescence. Due to the parity-forbidden nature of f-f transitions, the lanthanide-based UCPL phosphors show small absorption cross-sections, and NIR lasers with high power densities (eg. 7 W cm$^{-2}$) are needed to charge the phosphors. Later, Pan and coworkers realized the direct UCPL in $LiGa_5O_8$:Cr based on the excited state absorption and ionizable properties of $Cr^{3+}$.[18] Han and Qiu reported an upconversion-like $CaSnO_3$:$Bi^{2+}$ phosphor that can be directly charged by 700-740 nm light to produce persistent luminescence at 810 nm.[10] The NIR excitation is absorbed by the $^2P_{1/2} \rightarrow {}^2P_{3/2}$ transition of $Bi^{2+}$ (Figure 1b). Deep electron traps capture delocalized electrons from excited $Bi^{2+}$ and further pass the electrons to shallow traps. The de-trapping of electrons from shallow traps to $Bi^{3+}$ generates persistent luminescence. Albeit such achievements, ionizable activators are required to pass the photon energy to the conduction band (CB) or traps in the form of electrons, leading to significant limitations in the design of UCPL materials. The development of UCPL phosphors is currently seriously hindered by the limited photon upconversion processes that are possible to realize by the NIR light excitation.

Traps formed by lattice defects play center roles in the generation of persistent luminescence.[19, 20] Continuously distributed trap levels are generally present in the bandgap of persistent luminescence phosphors, which can function as middle transition levels for photon upconversion.[21] Moreover, the trap levels near the valence band maximum (VBM) are generally occupied by electrons.[22, 23] Therefore, we reason that the electrons in occupied trap levels or VB can be pumped to middle trap levels and further to high-energy trap levels or CB by NIR light. After NIR excitation ceases, the electrons trapped in high-energy trap levels migrate to activators *via* CB or quantum tunneling, leading to the generation of UCPL (Figure 1c). To validate our hypothesis, we chose $ZnGa_2O_4$:Cr (ZGO:Cr) nanoparticles with spinel configuration as the model since spinel features abundant vacancies and antisite defects. After pre-illumination by a broadband 800 nm LED, ZGO:Cr nanoparticles successfully display UCPL at about 700 nm for over 2 hours. Owing to the continuously distributed trap levels in the bandgap, ZGO:Cr nanoparticles can be directly charged by broadband NIR light in the range of 700-1000 nm. A trap-level-mediated photon upconversion mechanism is proposed for the NIR-to-NIR UCPL phenomenon in ZGO:Cr. The feasibility of the NIR-to-NIR UCPL for *in vivo* bioimaging has been verified, supporting the great potential of the developed UCPL phosphors in different imaging and display applications.

**Results and Discussion**

ZGO:Cr nanoparticles (Figure 2a and Figure S1) were prepared by a solvothermal reaction.[24,25] High-angle annular dark field scanning transmission electron microscopy (HAADF STEM) image and the corresponding elemental mapping images show the homogenous distribution of Zn, Ga, O, and Cr in ZGO:Cr (Figure 2b and 2c). To empty the possible photo-excited electrons trapped in ZGO:Cr during synthesis and storage, the nanoparticles were calcined at 700 °C for 1 hour before measurement (Figure S2a). Figure 2d shows that ZGO:Cr nanoparticles produce typical NIR emission of $Cr^{3+}$ at about 700 nm under the excitation of 850 nm or 800 nm light, demonstrating the successful NIR-to-NIR upconversion luminescence (UCL) in ZGO:Cr. Moreover, ZGO:Cr nanoparticles can be activated by NIR light in a broad wavelength range to generate upconversion luminescence (Figure S2b-2d). These results demonstrate the presence of broadband NIR photon upconversion process in ZGO:Cr, which indicates largely improved NIR absorption than the conventional lanthanide-based UCL nanoparticles.

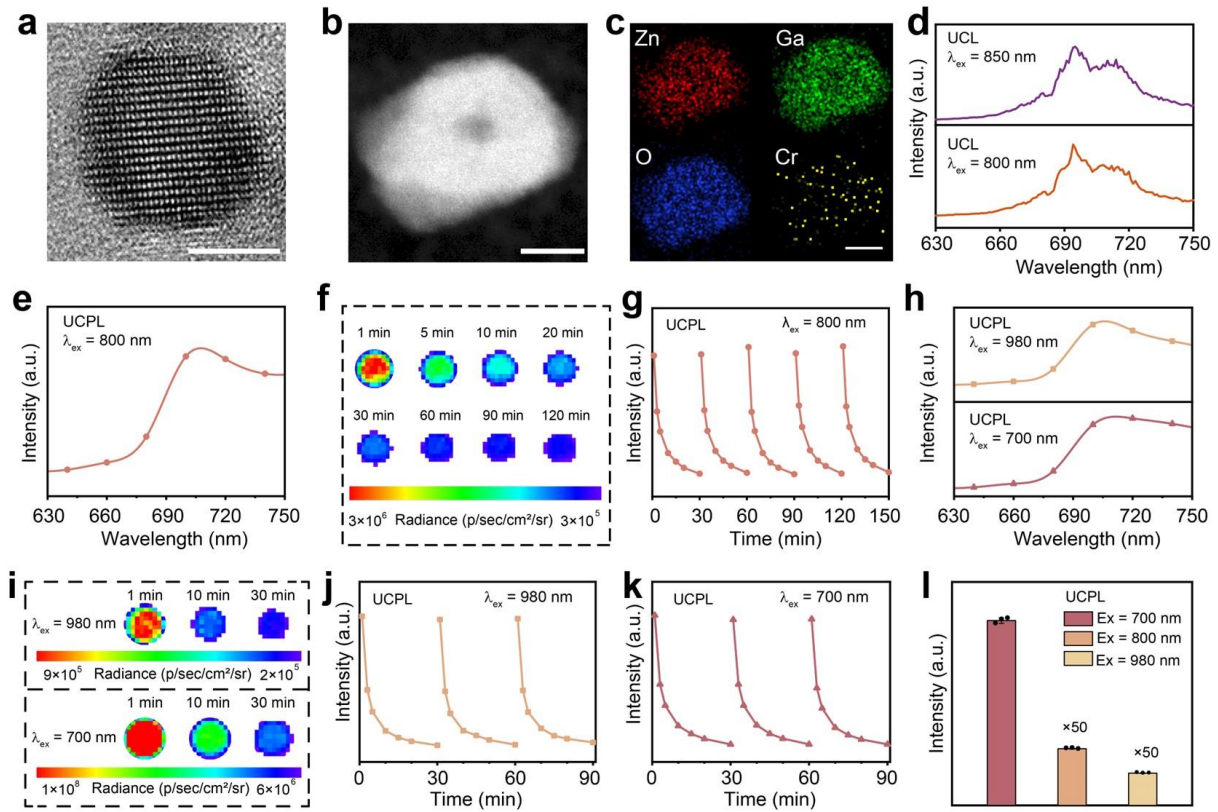

**Figure 2.** The broadband NIR-to-NIR UCL and UCPL in ZGO:Cr. (a) High-resolution TEM image of a ZGO:Cr nanoparticle. Scale bar, 5 nm. HAADF STEM image (b) and the corresponding elemental mapping images (c) of a ZGO:Cr nanoparticle. Scale bar, 5 nm. (d) UCL spectra of ZGO:Cr nanoparticles. (e) UCPL spectrum of ZGO:Cr nanoparticles after pre-excitation by an 800 nm LED. (f) UCPL decay images of ZGO:Cr nanoparticles. Weight, 50 mg. (g) Reproducibility of UCPL after repeated pre-excitation by an 800 nm LED. (h) UCPL spectra of ZGO:Cr nanoparticles after pre-excitation by 980 nm and 700 nm LED. (i) UCPL decay images of ZGO:Cr nanoparticles after pre-excitation by 980 nm and 700 nm LED. Reproducibility of UCPL after repeated pre-excitation by the 980 nm (j) and 700 nm (k) LED. (l) UCPL intensity of ZGO:Cr nanoparticles at 1min post-excitation by 700 nm, 800 nm and 980 nm LED.

The UCPL performances of ZGO:Cr nanoparticles are further investigated through illumination of a broadband 800 nm LED (Figure S3) for 5 min, where ZGO:Cr nanoparticles display typical NIR persistent luminescence at about 700 nm under ambient conditions (Figure 2e) and lasts for more than 2 hours (Figure 2f). This NIR persistent luminescence spectrum is also identical to that of ZGO:Cr pre-charged by 254 nm light (Figure S2e and S2f). Moreover, ZGO:Cr can be repeatedly charged by 800 nm light, without a decrease in UCPL intensity during the five reactivation cycles (Figure 2g and Figure S2g), which is remarkably different from the NIR light-stimulated persistent luminescence. The NIR light-stimulated persistent luminescence shows decreased luminescence intensity with increasing the number of NIR stimulations due to the gradual consumption of trapped charges.[26] The unchanged UCPL of ZGO:Cr in repeated reactivation (Figure 2g) demonstrates that ZGO:Cr can be directly charged by 800 nm light.[10] The NIR photon emission rate is employed to evaluate the UCPL intensity, where ZGO:Cr (50 mg) produces UCPL of about $1.29 \times 10^6$ p $sec^{-1}$ $sr^{-1}$ $cm^{-2}$ at 1 min post the 800 nm light excitation. Moreover, broadband NIR excitation at about 980 and 700 nm can both activate ZGO:Cr to produce NIR persistent luminescence (Figure 2h). The UCPL produced by 980 or 700 nm pre-excitation also remains after 2 hours of decay (Figure 2i, Figure S2h and S2i).

Similarly, ZGO:Cr can be repeatedly charged by 700 or 980 nm light without a decrease in UCPL intensity (Figure 2j, 2k, Figure S2j and S2k). The 700 nm broadband NIR excitation produces higher UCPL efficiency compared to 800 and 980 nm excitation (Figure 2l). These results reveal the successful NIR-to-NIR UCPL in ZGO:Cr nanoparticles, which can be directly and repeatedly charged by broadband NIR light in the range of 700-1000 nm without affecting the UCPL intensity. The remarkable UCPL performances are ascribed to the continuously distributed intermediate trap levels in the bandgap.

We attributed the NIR-to-NIR UCPL in ZGO:Cr to the ladder-like trap levels in the bandgap. Thus, elevating the density of trap levels in ZGO:Cr should favor the UCPL process. As a proof of concept experiment, we investigated whether the UCPL of ZGO:Cr can be enhanced by $Ge^{4+}$ or $Sn^{4+}$ doping.[24] $Zn_{1.2}Ga_{1.6}Ge_{0.2}O_4$:Cr (ZGGO:Cr) and $Zn_{1.1}Ga_{1.8}Sn_{0.1}O_4$:Cr (ZGSO:Cr) with elevated trap density exhibit typical UCL upon the excitation of NIR light (Figure 3a, 3b, Figure S5c-S5f and Figure S7c-S7f). The NIR persistent luminescence that lasts for more than 2 h is detected in ZGGO:Cr and ZGSO:Cr nanoparticles after pre-charging by broadband NIR light (Figure 3c-3e, Figure S5g-S5k and Figure S7i). Similarly, ZGGO:Cr and ZGSO:Cr can be directly charged by NIR light and the persistent luminescence decay remains unchanged in the repeated activation cycles (Figure S5i- S5n and Figure S7j). The UCPL intensity in ZGGO:Cr and ZGSO:Cr was further quantified by measuring the NIR photon emission rate. At 1 min post-illumination by 700 nm light, the persistent luminescence intensity of ZGGO:Cr and ZGSO:Cr is about $3.97 \times 10^8$ p sec$^{-1}$ sr$^{-1}$ cm$^{-2}$ and $4.01 \times 10^8$ p sec$^{-1}$ sr$^{-1}$ cm$^{-2}$, 1.92 and 1.94 times that of ZGO:Cr, respectively (Figure 3f). We noticed that the size of ZGGO:Cr (22 nm, Figure S5b) and ZGSO:Cr (15 nm, Figure S7b) is larger than that of ZGO:Cr (10 nm, Figure S1c) after calcination. The enhancement of UCPL intensity in ZGGO:Cr and ZGSO:Cr may be due to their increased size.[27] To confirm that the elevated density of trap level contributes to the enhanced UCPL, we prepared ZGO:Cr nanoparticles with a larger size (L-ZGO:Cr, 24 nm) than that of ZGGO:Cr and ZGSO:Cr based on a homogeneous epitaxial shell growth strategy (Figure S8 and S9).[28] The L-ZGO:Cr nanoparticles produce UCPL stronger than ZGO:Cr after NIR excitation ceases, but much weaker than that of ZGGO:Cr and ZGSO:Cr (Figure S9d). The stronger persistent luminescence from Ge/Sn-doped ZGO:Cr suggests that the UCPL is associated with trap levels.

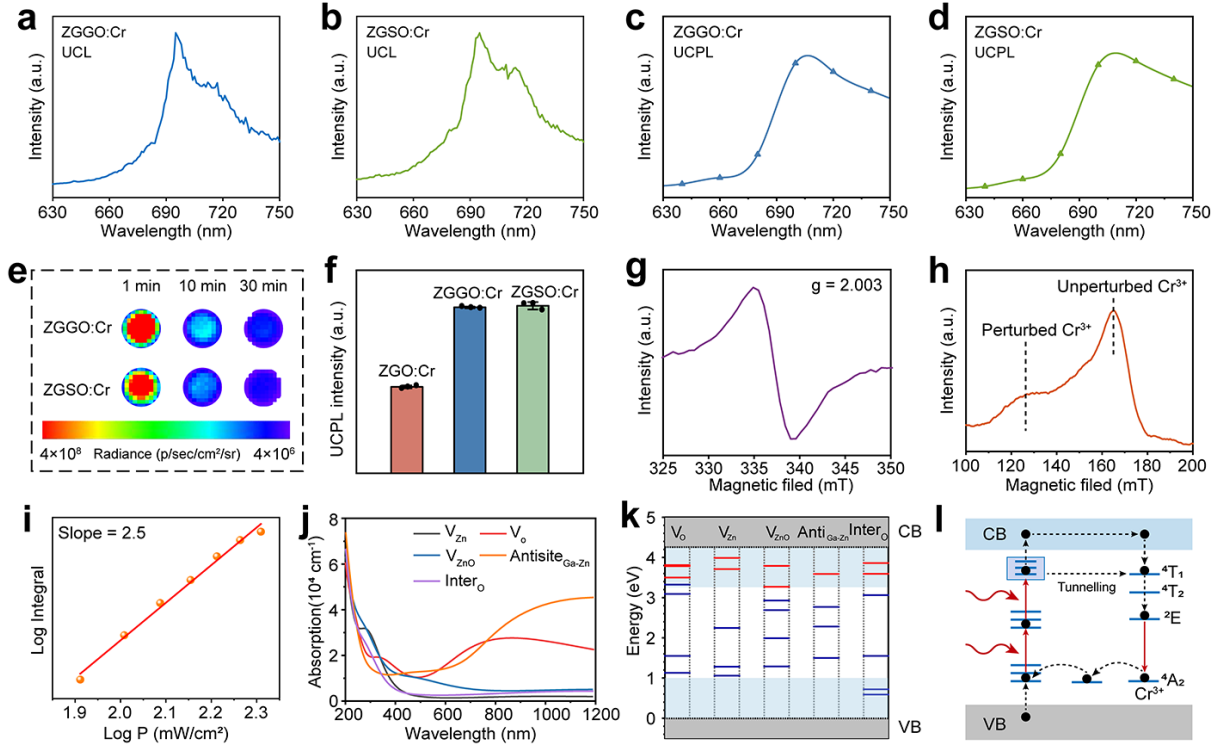

**Figure 3.** Mechanism of the NIR-to-NIR UCPL in ZGO:Cr. UCL spectra of ZGGO:Cr (a) and ZGSO:Cr (b) nanoparticles. Excitation, 800 nm. UCPL spectra (c, d) and UCPL decay images (e) of ZGGO:Cr and ZGSO:Cr nanoparticles. Excitation, 800 nm. (f) UCPL intensity of ZGO:Cr, ZGGO:Cr and ZGSO:Cr nanoparticles at 1min post-excitation. EPR spectroscopy of the oxygen vacancy (g) and $Cr^{3+}$ (h) in ZGO:Cr nanoparticles. (i) The double-logarithmic plot of UCL intensity versus the power of an 800 nm laser. (j) The simulated absorption spectra of different defects. (k) The summarized trap levels induced by different defects in ZGO:Cr. The red lines represent the empty states, the blue lines represent the occupied states. The blue shadow indicates the 1 eV trap depth from CBM and VBM. (l) Schematic illustration of the two-photon UCPL process in ZGO:Cr.

We further investigated the possible NIR-to-NIR UCPL mechanism in ZGO:Cr nanoparticles. Our previous Rietveld refinement assays confirmed the presence of antisite defects in ZGO.[29] Electron paramagnetic resonance (EPR) spectroscopy and thermoluminescence measurements were used to probe the defective features of ZGO:Cr. Figure 3g suggests the presence of oxygen vacancy ($V_O$) occupied by an electron in ZGO:Cr nanoparticles.[30] The EPR spectrum of $Cr^{3+}$ in ZGO:Cr shows that considerable amounts of $Cr^{3+}$ ions locate at octahedral sites distorted by neighboring charged defects (Figure 3h),[31] reassuring the presence of abundant lattice defects in ZGO:Cr. Furthermore, the thermoluminescence spectrum of ZGO:Cr spans from 20 to 150 °C (Figure S10a), which signifies the presence of continuously distributed trap levels formed by lattice defects.[32] We next investigated the trap-level-mediated photon upconversion process by recording the UCL intensity (I) of ZGO:Cr versus the power density (P) of an 800 nm laser. The UCL intensity increases upon elevating the power density of the laser (Figure S10b). The linear regression fits on the double-logarithmic plot of UCL intensity versus the power of an 800 nm laser gives a slope value of 2.5 (Figure 3i), suggesting the presence of two-photon and three-photon upconversion processes in ZGO:Cr.[33,34]

To further gain fundamental insights into the UCPL mechanism, we performed first-principles density functional theory (DFT) investigations on the trap levels originating from lattice defects (Figure S11). Zn vacancy($V_{Zn}$), Schotty-like ZnO vacancy ($V_{ZnO}$), interstitial oxygen ($Inter_O$), and $Antisite_{Ga-Zn}$ display spontaneous formation trends in ZGO (Figure S11d), supporting the abundant defects in ZGO lattices during the synthesis. According to the optical absorption of ZGO:Cr with these defects, the $V_O$ and $Antisite_{Ga-Zn}$ exhibit significant improvements to the absorption intensities of ZGO:Cr from 600 to 1200 nm (Figure 3j). Schotty-like $V_{ZnO}$ also enhances the absorption intensities in 400-1200 nm, suggesting that $V_O$, $V_{ZnO}$, and $Antisite_{Ga-Zn}$ in ZGO:Cr play significant roles in absorbing the 700-1000 nm NIR excitation. Based on the projected partial density of states (PDOS) of different defects, we further summarized the trap level distributions in ZGO:Cr (Figure 3k). Different defects induce abundant trap levels with different depths in the bandgap, with empty trap levels mainly located near CBM (red line) and occupied trap levels widely distributed in the bandgap (blue line). We noticed that the energy gap between the adjacent trap levels is below 1.8 eV, suggesting that light with wavelengths longer than 690 nm is sufficient to pump the transition of electrons between these adjacent trap levels. In light of experimental and theoretical results, we proposed a trap-level-mediated upconversion charging mechanism to interpret the NIR-to-NIR UCPL in ZGO:Cr (Figure 3l). Upon broadband NIR excitation, the electrons in intermediate occupied trap levels absorb NIR photons and are pumped to high energy empty trap levels or CB. Meanwhile, the electrons in low-energy occupied trap levels or VB transfer to the intermediate trap levels upon absorbing NIR photons. In this way, the electrons in occupied trap levels or VB are pumped to empty trap levels beneath CBM for energy storage through such a two-photon upconversion process. Notably, the intermediate occupied trap levels are capable of stabilizing electrons, which is beneficial for the NIR photon upconversion process. After excitation ceases, the trapped electrons escape to the conduction band after absorbing the heat available at room temperature, followed by the migration and annihilation of electrons at the $Cr^{3+}$ sites to produce NIR persistent luminescence.

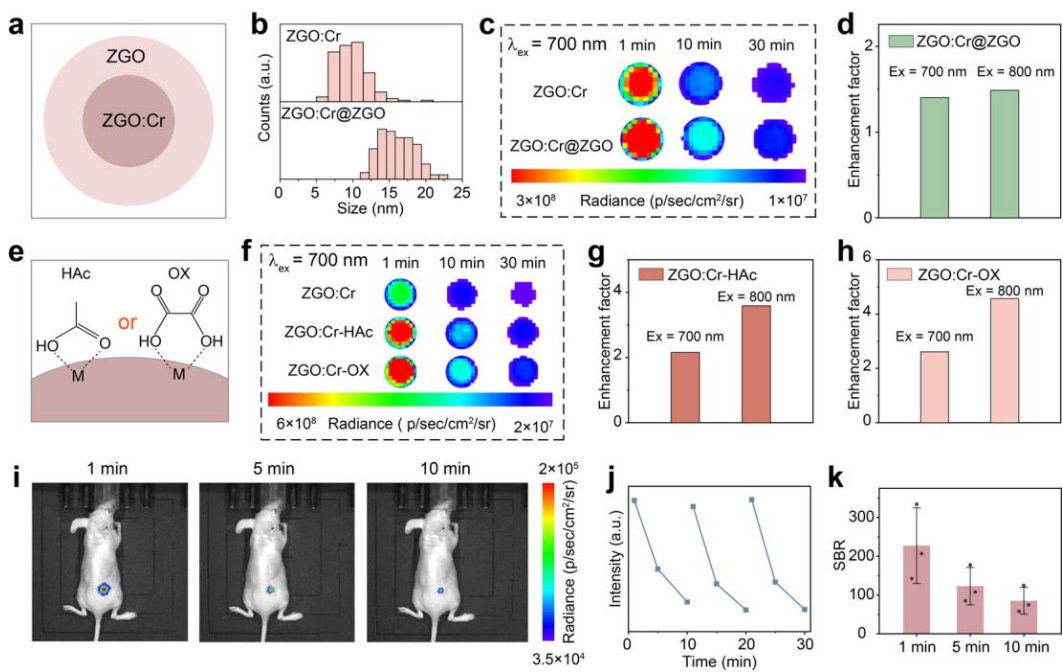

**Figure 4.** Enhancing UCPL in ZGO:Cr by surface passivation and exploring the bioimaging potential of ZGO:Cr. (a) Schematic illustration of surface passivation by ZGO shell. The size distribution (b) and UCPL decay images (c) of ZGO:Cr and ZGO:Cr@ZGO nanoparticles. (d) UCPL enhancement factor of ZGO:Cr@ZGO compared to ZGO:Cr. (e) Schematic illustration of surface passivation by HAc and OX. (f) UCPL decay images of ZGO:Cr, ZGO:Cr-HAc and ZGO:Cr-OX nanoparticles. UCPL enhancement factor of ZGO:Cr-HAc (g) and ZGO:Cr-OX (h) compared to ZGO:Cr. (i) UCPL images of mice bearing subcutaneously injected ZGGO:Cr after ceasing the 700 nm LED. (j) UCPL intensities of subcutaneously injected ZGGO:Cr upon repeated *in vivo* activation. (k) The SBR of the subcutaneously injected ZGGO:Cr as a function of time.

We have shown elevating the density of trap levels in ZGO by Ge/Sn doping to improve the UCPL efficiency. Considering that the luminescence of nanoparticles is strongly attenuated by surface quenching,[2,35,36] we next explored enhancing the NIR-to-NIR UCPL in ZGO by surface passivation. Core-shell structured ZGO:Cr@ZGO nanoparticles were prepared by a seed-mediated growth strategy (Figure 4a and Figure S13a).[28] A ZGO shell of about 3 nm is coated on the surface of ZGO:Cr (Figure 4b). The ZGO:Cr@ZGO nanoparticles produce typical NIR persistent luminescence that lasts for over 2 h after NIR excitation (Figure 4c and Figure S13b). The UCPL intensity in ZGO:Cr@ZGO at 1 min post excitation by 700 and 800 nm light is about 1.4 and 1.5 times of bare ZGO:Cr, respectively (Figure 4d), which is ascribed to the passivation of surface quenching sites by a ZGO shell. Later, ligands acetic acids (HAc) and oxalate (OX) with coordination ability were used to passivate the surface defects on ZGO:Cr nanoparticles to enhance their UCPL (Figure 4e). Both ZGO:Cr-HAc and ZGO:Cr-OX nanoparticles produce durable persistent luminescence after excitation by 700 or 800 nm light (Figure 4f and Figure S13-S15). The UCPL of surface coordinated ZGO:Cr nanoparticles is more than two times stronger than that of bare ZGO:Cr (Figure 4g and 4h). ZGO:Cr-OX nanoparticles produce stronger UCPL than ZGO:Cr-HAc, which can be attributed to the formation of a stable five-membered ring between bidentate OX and the surface metal ions of ZGO:Cr.[35]

Finally, we tested the potential applications of the NIR-to-NIR UCPL in bioimaging. ZGGO:Cr nanoparticles were excited by 700 nm LED either with or without the coverage of pork tissue of different thicknesses (1-2 cm), and the UCPL signal passing through the pork tissue was monitored (Figure S16a). ZGGO:Cr charged in the presence of pork coverage produces UCPL intensity comparable to the directly charged ZGGO:Cr (Figure S16b), suggesting the effectiveness of the *in vivo* activation of ZGGO:Cr by NIR light. The ZGGO:Cr nanoparticles (4 mg/mL, 25 μL) were further subcutaneously injected into mice for further bioimaging testing. After *in vivo* activation by 700 nm LED, the ZGGO:Cr displays reliable photostability and reproducibility upon *in vivo* reactivation (Figure 4i and Figure S16c), with an unaltered initial intensity and decay duration (Figure 4j). The signal-to-background ratio (SBR) is determined to be about 230 at 1 min post excitation (Figure 4k), which outperforms the previously reported UCPL phosphors. Although the UCPL intensity decays over time, the SBR is kept over 85 even at 10 min post excitation. The high SBR is ascribed to the elimination of tissue autofluorescence and light scattering interference.

**Conclusion**

This work reported a new photon upconversion phenomenon mediated by lattice defects. The broadband NIR absorption of ZGO:Cr enhances the efficiency of photon upconversion, which is advantageous over UCPL based on the narrow-band NIR absorption of lanthanides.[37] Our discovery elucidates the important roles of occupied trap levels in the generation of persistent luminescence, which have been previously ignored in most research. Notably, defect density, distribution of trap levels in the bandgap, and the bandgap of phosphors are critical in the UCPL process. It is worth noting that a suitable defect density is highly significant for the UCPL. On one side, elevating the defect density enhances the NIR light absorption capacity and electron trapping capacity of phosphors, which promotes the UCPL process. On the other side, excessive defects quench the UCPL intensity (Figure S17) since trap levels can transfer the excitation energy to surface quenching sites.[38] Accordingly, the balance of the defect density is significant to achieve the maximum UCPL efficiency. In addition to defect intensity, the distribution of trap levels also plays a crucial role in UCPL. The continuously distributed trap levels in the bandgap are essential for NIR photon upconversion. In the end, the bandgap of phosphors influences UCPL efficiency. Since the electrons in VB or the occupied trap levels need to be pumped to CB or empty trap levels beneath CB, phosphors with large bandgap (eg. > 6 eV)[21] require three or four-photon upconversion processes to charge the phosphors, where multi-photon upconversion inevitably decreases UCPL efficiency. Based on our work, the defect density, distribution of trap levels, and bandgap of phosphors need to be collectively optimized to achieve strong and durable UCPL.

To conclude, we have discovered a new NIR-to-NIR UCPL in ZGO:Cr nanoparticles and proposed a trap-level-mediated broadband upconversion mechanism to interpret the UCPL process, where the lattice defects are critical in the generation of UCPL. Depending on the assistance of defects, the electrons in occupied trap levels or VB can be directly pumped to intermediate trap levels and further to empty trap levels beneath CBM by broadband NIR light. The de-trapping of the abundant trapped electrons to $Cr^{3+}$ after excitation ceases effectively produces NIR persistent luminescence. In particular, the UCPL process is realized under the broadband NIR excitation in the range of 700-1000 nm due to the continuously distributed trap levels in the bandgap of ZGO:Cr. In addition, ion doping and surface passivation have been proven to be effective approaches to enhance the UCPL in ZGO:Cr nanoparticles, where the ZGGO:Cr nanoparticles achieved an SBR of up to 230 in bioimaging. Our work reported a novel defect-assisted mechanism for the design of UCPL, which opens new opportunities for the rational design of UCPL phosphors towards applications including bioimaging, phototheranostics, photocatalysis, and solar-to-chemical synthesis.

**Experimental Methods**

Chemical reagents. Zinc nitrate hexahydrate ($Zn(NO_3)_2·6H_2O$, 99%), gallium oxide ($Ga_2O_3$, 99.99%), germanium oxide ($GeO_2$, 99.99%), chromium nitrate nonahydrate ($Cr(NO_3)_3·9H_2O$), sodium stannate trihydrate ($Na_2SnO_3·3H_2O$, 98%) and ammonium hydroxide ($NH_3·H_2O$, 25 wt%) were purchased from Aladdin (Shanghai, China). Acetic acid (HAc, 99.5%) and oxalic acid (OX, 99.5%) were purchased from Aladdin (Shanghai, China). Concentrated nitric acid ($HNO_3$) and sodium hydroxide (NaOH) were purchased from Sinopharm Chemical Reagent Co. (Shanghai, China). $Ga_2O_3$ was added to diluted $HNO_3$ and reacted overnight at 120 °C to obtain $Ga(NO_3)_3$ solution. $Na_2GeO_3$ solution was obtained by dissolving $GeO_2$ in NaOH solution. Typically, 1 M $Zn(NO_3)_2$, 0.4 M $Ga(NO_3)_3$, 0.4 M $Na_2GeO_3$, 0.4 M $Na_2SnO_3$ and 0.08 M $Cr(NO_3)_3$ were stored as precursor solutions in this experiments.

Characterization. The sizes and shapes of nanoparticles were determined by a HT-7700 transmission electron microscope (TEM, HITACHI, Japan). The high-resolution TEM (HRTEM) was performed on a TEM (Thermo Scientific, Talos F200X G2, Czech). The high-angle annular dark-field scanning transmission electron microscopy (HAADF-STEM) and elemental mappings were conducted on a Themis Z TEM (Thermo Scientific, Netherlands) with an accelerating voltage of 300 kV. Powder X-ray diffraction (XRD) characterization was conducted at room temperature on a Bruker D8 diffractometer. Persistent luminescence decay images and intensities were measured by an IVIS imaging system (Perkin-Elmer, UK). Upconversion luminescence spectra were measured with the FLS980 spectrometer (Edinburgh) equipped with a tunable mid-band optical parametric oscillator (OPO) pulse laser as the excitation source (410-2400 nm, 10 Hz, pulse width 5 ns, Vibrant 355ll, Opotek). Phosphorescence spectra of nanoparticles were measured on a fluorescence spectrometer (Hitachi, F-4600, Japan). Electron paramagnetic resonance (EPR) characterization was performed on an ESR spectrometer (Bruker, A300,

Germany) at low temperature. The thermoluminescence (TL) assay was carried out with a TL spectroscopy (TOSL-3DS, China). Absorption spectra of nanoparticles were collected with a UV-vis-NIR spectrophotometer (Shimadzu, UV-3600, Japan). Emission spectra of NIR LED were measured on a fluorescence spectrometer (Avantes, AvaSpec-2048).

Synthesis of ZGO:Cr nanoparticles. Typically, 1 mmol $Zn(NO_3)_2$, 2 mmol $Ga(NO_3)_3$ and 0.01 mmol $Cr(NO_3)_3$ were added to 11 mL deionized water. Next, concentrated $NH_3·H_2O$ was quickly added to the above solution under vigorous stirring to adjust the pH of the solution to around 8.5, during which white precipitates appeared. Then, the mixture was left stirring for 1 h at room temperature. After that, the solution was transferred to a Teflon-lined autoclave and heated at 220 °C for 10 h. The obtained nanoparticles were washed three times with deionized water and dried at 80 °C.

Synthesis of $Zn_{1+x}Ga_{2-2x}Ge_xO_4$:Cr (x = 0.1-0.3) nanoparticles. The synthesis of ZGGO:Cr nanoparticles with the composition of x = 0.2 is used as an example. Typically, 1.2 mmol $Zn(NO_3)_2$, 1.6 mmol $Ga(NO_3)_3$, 0.2 mmol $Na_2GeO_3$ and 0.01 mmol $Cr(NO_3)_3$ were added to 11 mL deionized water. The subsequent steps were the same as the synthesis of ZGO:Cr.

Synthesis of ZGSO:Cr. Briefly, 1.1 mmol $Zn(NO_3)_2$, 1.8 mmol $Ga(NO_3)_3$, 0.1 mmol $Na_2SnO_3$ and 0.01 mmol $Cr(NO_3)_3$ were added to 11 mL deionized water. The subsequent steps were the same as synthesis of ZGO:Cr.

Synthesis of L-ZGO:Cr nanoparticles. Briefly, 0.1 g ZGO:Cr, 0.5 mmol $Zn(NO_3)_2$, 1 mmol $Ga(NO_3)_3$ and 0.005 mmol $Cr(NO_3)_3$ were added to 11 mL deionized water. The subsequent steps were the same as the synthesis of ZGO:Cr.

Synthesis of ZGO:Cr@ZGO nanoparticles. Typically, 0.1 g ZGO:Cr, 0.5 mmol Zn(NO$_3$)$_2$ and 1 mmol Ga(NO$_3$)$_3$ were added to 11 mL deionized water. The subsequent steps were the same as the synthesis of ZGO:Cr.

Synthesis of ZGO:Cr-HAc and ZGO:Cr-OX nanoparticles. Typically, 1 mmol Zn(NO$_3$)$_2$, 2 mmol Ga(NO$_3$)$_3$, 0.01 mmol Cr(NO$_3$)$_3$ and 0.25 mmol HAc or OX were added to 11 mL deionized water. The subsequent steps were the same as the synthesis of ZGO:Cr.

UCPL tests. The nanoparticles were calcined at 700 °C for 1 hour to empty the possible photo-excited electrons trapped in lattice defects during synthesis and storage. Briefly, nanoparticles (50 mg) were put into a 96-plate well. The nanoparticles were illuminated by different LEDs (700 or 800 nm for 10 min, 980 nm for 2 min). After illumination, the 96-plate well was placed in an IVIS Lumina XR III imaging system to capture the persistent luminescence decay images. The exposure time was set as 20 s for 800 nm light-charged nanoparticles, 1 s for 700 nm light-charged nanoparticles, and 40 s for 980 nm light-charged nanoparticles.

Theoretical calculations. Theoretical calculations based on DFT are performed by CASTEP packages to investigate electronic modulations induced by the defects in ZGO.[39] The generalized gradient approximation (GGA) and Perdew-Burke-Ernzerhof (PBE) are applied in this work, which can offer accurate descriptions of the exchange-correlation interactions.[40-42] With the ultrasoft pseudopotentials, the plane-wave basis cutoff energy has been generated as 380 eV with ultrafine quality. The Broyden-Fletcher-Goldfarb-Shannon (BFGS) algorithm is applied for the energy minimizations, where the k-point settings have been set to 2 × 2 × 2 with coarse quality.[43] Moreover, we have applied the following convergence criteria including 1) the Hellmann-Feynman forces should not exceed 0.001 eV/Å; 2) the total energy difference should not be over 5×10$^{-5}$ eV/atom; and 3) the inter-ionic displacement should be less than 0.005 Å.

*In vivo* imaging tests. All animal experiments were carried out according to the protocols approved by Soochow University Laboratory Animal Center (approval number 202309A0367). BALB/c mice (6-8 weeks, female, 20 g) were purchased from the Laboratory Animal Center of Suzhou Medical College, Soochow University. Mice were anesthetized by isoflurane and subcutaneously injected with ZGGO: Cr aqueous solution (25 μL, 4 mg/mL). After being excited by a 700 nm LED for 5 min, the mice were put into an IVIS Lumina XRMS Series III imaging system for UCPL signal acquisition.

ASSOCIATED CONTENT

**Supporting Information**.

The following files are available free of charge.

TEM images and XRD patterns, photoluminescence and persistent luminescence images of ZGO:Cr NPs, and UCPL in ZGO:Cr NPs and bioimaging images (PDF)

AUTHOR INFORMATION

**Corresponding Author**

*bhuang@polyu.edu.hk. *jiewang@suda.edu.cn.

**Author Contributions**

‡These authors contributed equally.

**Funding Sources**

The authors acknowledge support from the National Natural Science Foundation of China (grant no. 22274105 and 21904100), the Gusu Innovation and Entrepreneurship Leading Talent Plan (ZXL2023199), the National Key R&D Program of China (2021YFA1501101), the National Natural Science Foundation of China/Research Grant Council of Hong Kong Joint Research Scheme (N_PolyU502/21), National Natural Science Foundation of China/Research Grants Council of Hong Kong Collaborative Research Scheme (CRS_PolyU504_22).

**Notes**

There is no conflict of interest to report.

ACKNOWLEDGMENT

The authors thank the support from the Research Centre for Carbon-Strategic Catalysis (RC-CSC), Research Institute for Smart Energy (RISE), and Research Institute for Intelligent Wearable Systems (RI-IWEAR) of the Hong Kong Polytechnic University. The authors sincerely thank Prof. Liangliang Liang from Xiamen University for his helpful suggestions that greatly improved the quality of the manuscript.

ABBREVIATIONS

UCPL, upconversion persistent luminescence; NIR, near-infrared; CB, conduction band; VBM, valence band maximum; ZGO:Cr, $ZnGa_2O_4$:Cr; HAADF STEM, High-angle annular dark field scanning transmission electron microscopy; UCL, upconversion luminescence; ZGGO:Cr, $Zn_{1.2}Ga_{1.6}Ge_{0.2}O_4$:Cr; ZGSO:Cr, $Zn_{1.1}Ga_{1.8}Sn_{0.1}O_4$:Cr; L-ZGO:Cr, $ZnGa_2O_4$:Cr@$ZnGa_2O_4$:Cr; EPR, Electron paramagnetic resonance; I, intensity; P, power density; DFT, density functional theory; $V_{Zn}$, Zn vacancy; $V_{ZnO}$, ZnO vacancy; $Inter_O$, interstitial oxygen; PDOS, partial density of states; HAc, acetic acids; OX, oxalate.

Table of Contents

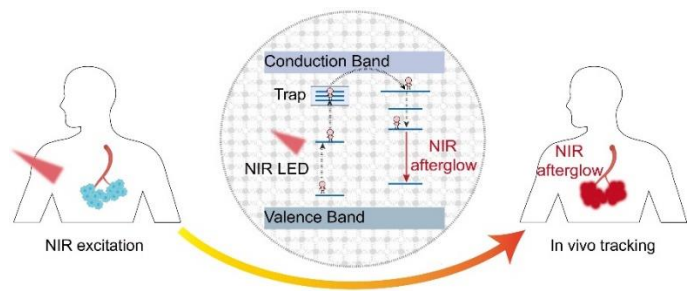

We discovered a crystal defect-mediated NIR photon upconversion phenomenon and the accompanied NIR persistent luminescence. The upconversion persistent luminescence nanoparticles can be readily reactivated *in vivo* by broadband NIR light to produce NIR afterglow, which makes the nanoparticles valuable in deep-tissue bioimaging and *in vivo* tracking.

# Broadband NIR photon upconversion generates NIR persistent luminescence for bioimaging

Shuting Yang,[1,‡] Bing Qi,[1,‡] Mingzi Sun,[2] Wenjing Dai,[1] Ziyun Miao,[1] Wei Zheng,[3] Bolong Huang,[2,*] Jie Wang[1,*]

[1] The Key Lab of Health Chemistry & Molecular Diagnosis of Suzhou, College of Chemistry, Chemical Engineering & Materials Science, Soochow University, Suzhou 215123, China.

[2] Department of Applied Biology and Chemical Technology, The Hong Kong Polytechnic University, Hung Hom, Kowloon, Hong Kong SAR, China.

[3] CAS Key Laboratory of Design and Assembly of Functional Nanostructures, Fujian Key Laboratory of Nanomaterials, and State Key Laboratory of Structural Chemistry, Fujian Institute of Research on the Structure of Matter, Chinese Academy of Sciences, Fuzhou, Fujian 350002, China.

[‡] These authors contributed equally to this work.

[*] Bolong Huang: bhuang@polyu.edu.hk; Jie Wang: jiewang@suda.edu.cn.

## Table of Contents

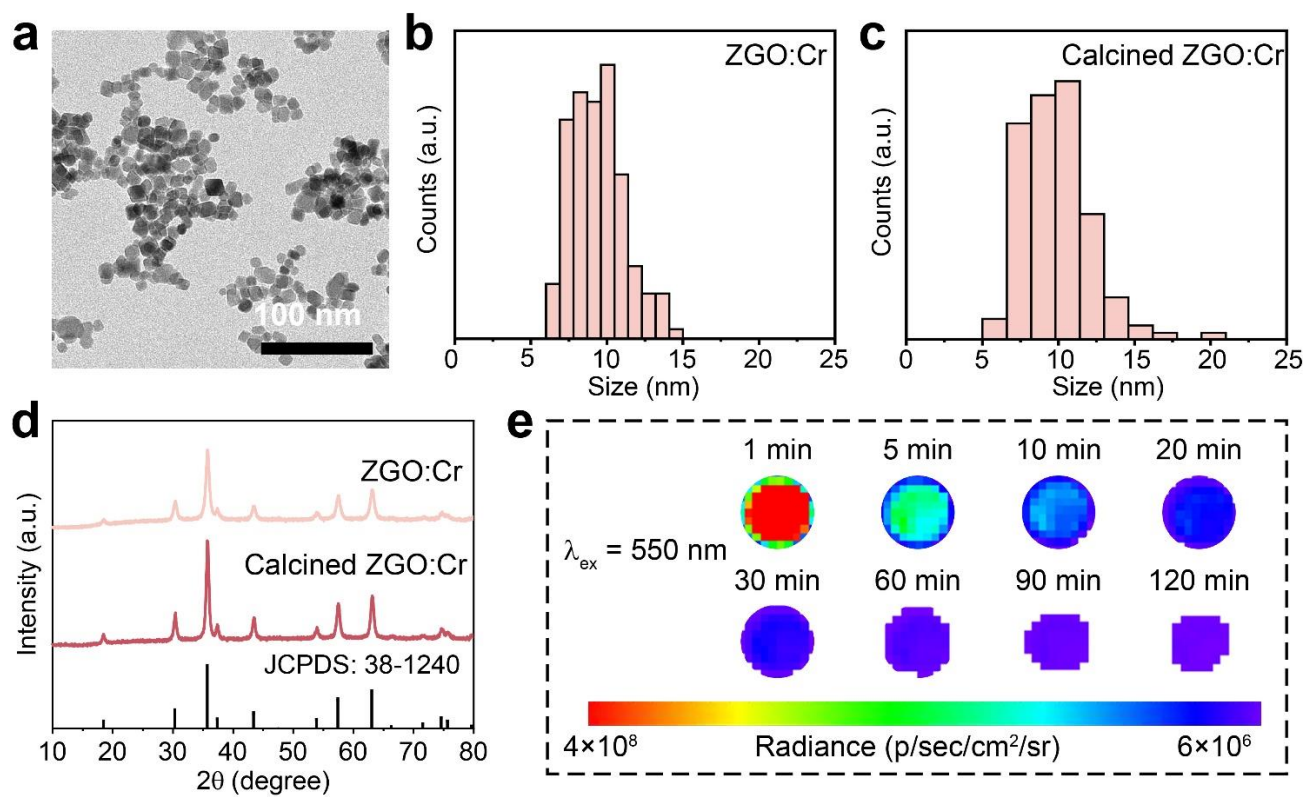

**Figure S1 Characterization of ZGO:Cr nanoparticles. a**, TEM image of ZGO:Cr nanoparticles. Size distributions of ZGO:Cr nanoparticles before (**b**) and after (**c**) calcination. **d**, X-ray diffraction (XRD) patterns of ZGO:Cr nanoparticles before and after calcination. **e**, Persistent luminescence decay images of ZGO:Cr nanoparticles after pre-excitation by 550 nm LED.

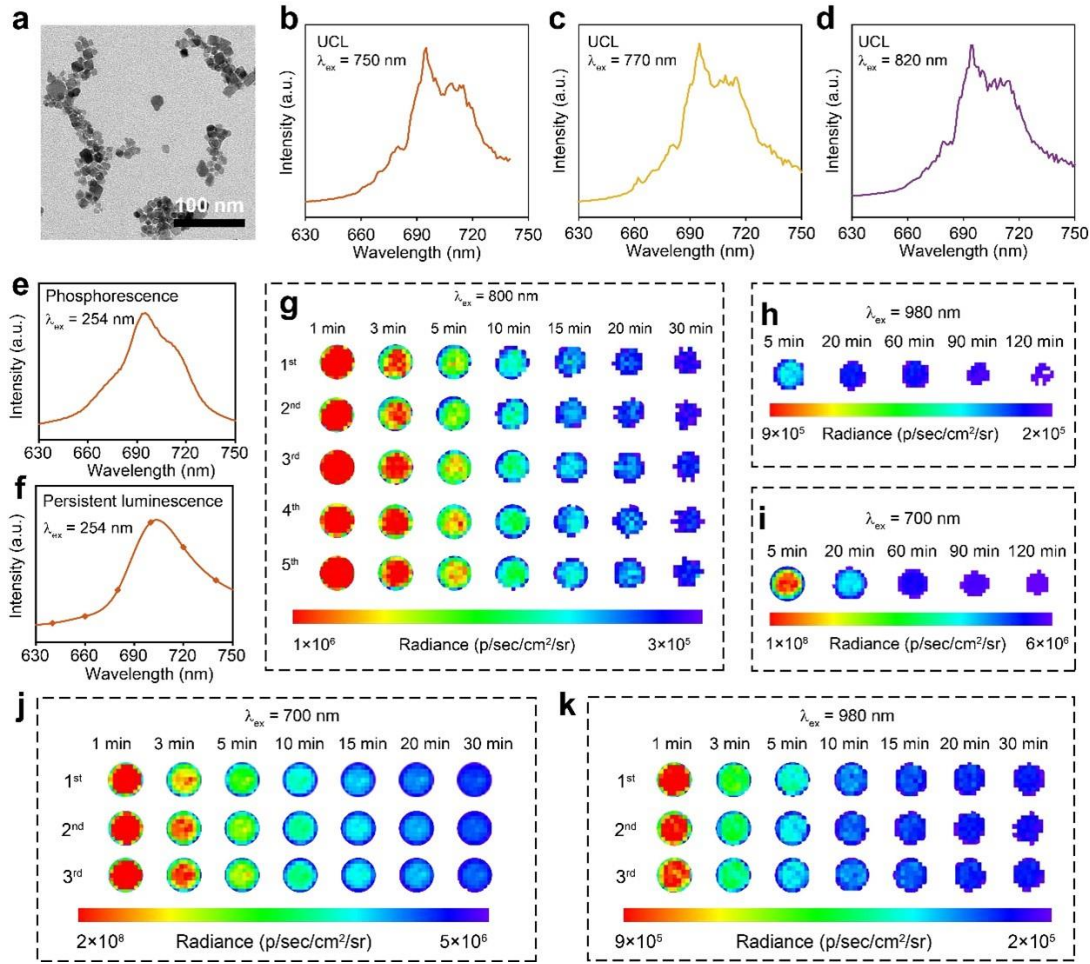

**Figure S2 UCPL in ZGO:Cr nanoparticles. a**, TEM image of ZGO:Cr nanoparticles after calcination. **b-d**, UCL spectra of ZGO:Cr nanoparticles under NIR excitation under different wavelengths. Phosphorescence (**e**) and persistent luminescence (**f**) spectrum of ZGO:Cr nanoparticles under 254 nm excitation. UCPL decay images of ZGO:Cr nanoparticles after pre-excitation by 700 nm (**i**) and 980 nm (**h**) LED. UCPL decay images of ZGO:Cr nanoparticles upon repeated pre-excitation by 800 nm (**g**), 700 nm (**j**), and 980 nm (**k**) LED.

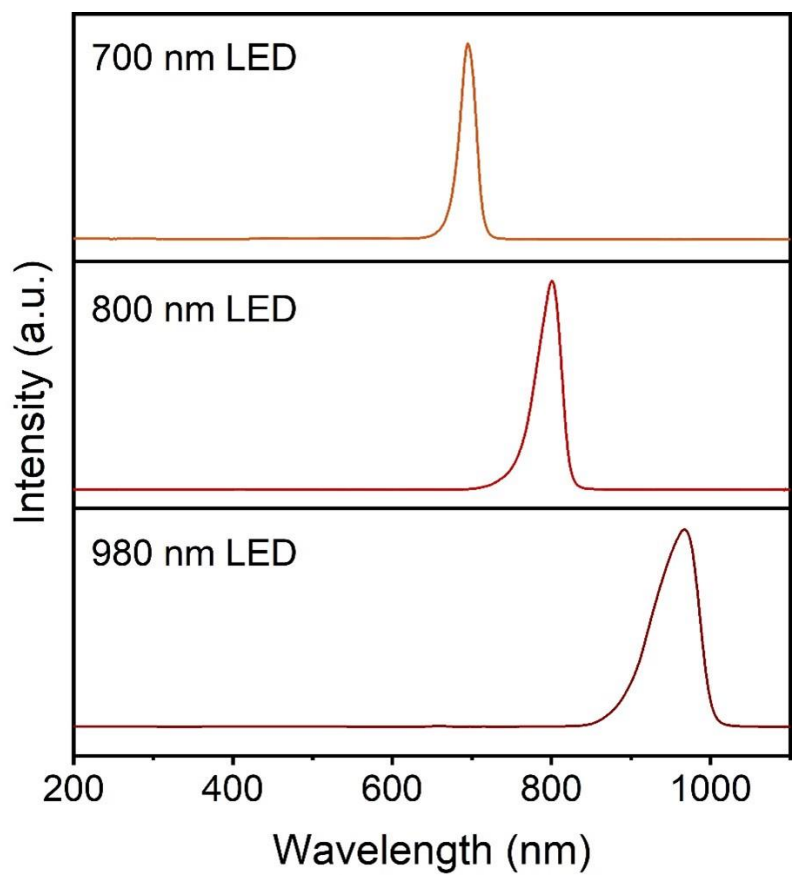

**Figure S3 Emission spectra of the 700 nm, 800 nm, and 980 nm LED used in this study.**

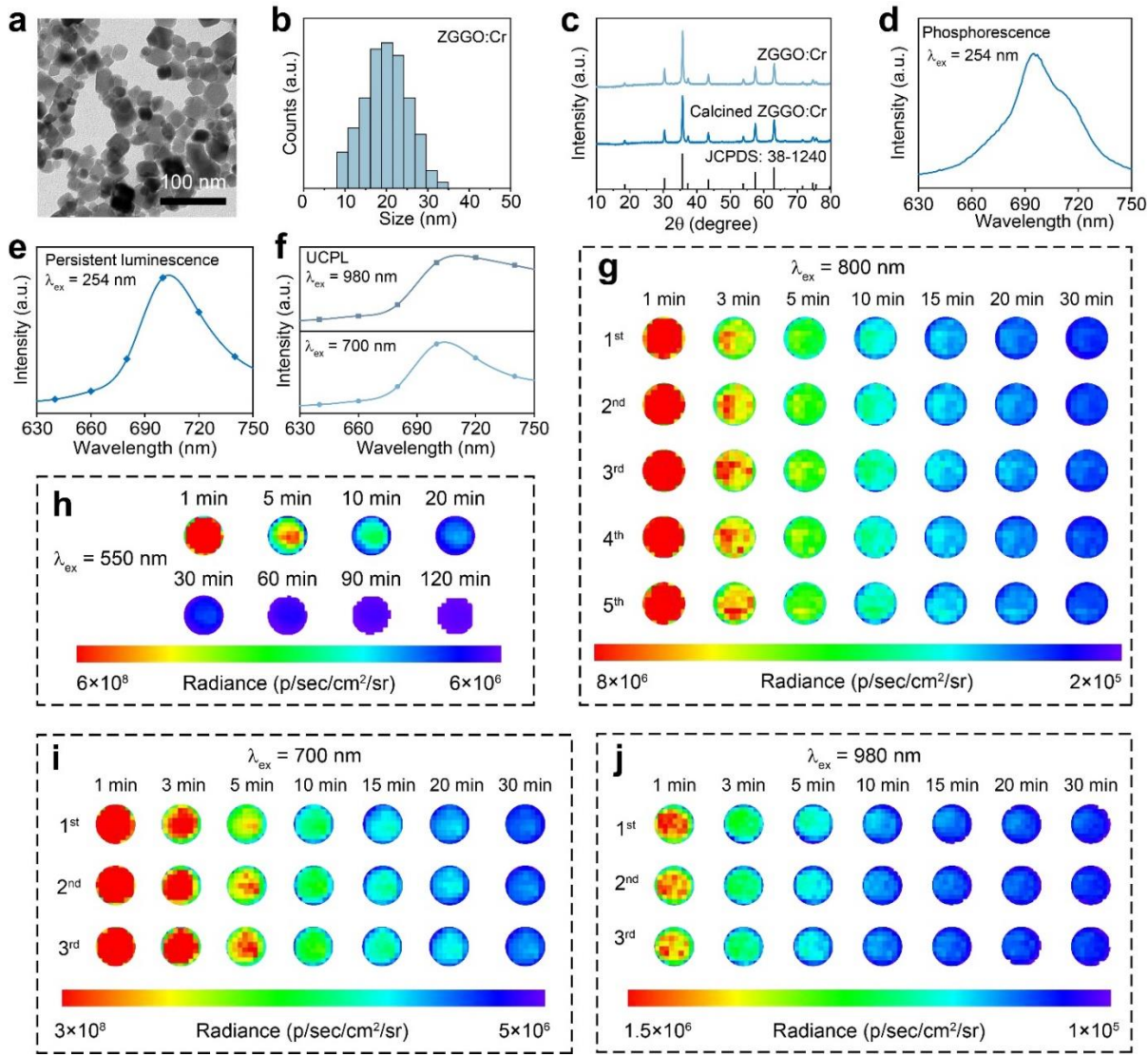

**Figure S4 Characterization of ZGGO:Cr nanoparticles.** TEM image (**a**) and the size distribution (**b**) of ZGGO:Cr nanoparticles before calcination. **c**, XRD patterns of ZGGO:Cr nanoparticles before and after calcination. Phosphorescence (**d**) and persistent luminescence (**e**) spectrum of ZGGO:Cr nanoparticles under 254 nm excitation. **f**, UCPL spectra of ZGGO:Cr nanoparticles after pre-excitation by different NIR LED. **g**, UCPL decay images of ZGGO:Cr nanoparticles upon repeated excitation by 800 nm LED. **h**, Persistent luminescence decay images of ZGGO:Cr nanoparticles after pre-excitation by 550 nm LED. UCPL decay images of ZGGO:Cr nanoparticles upon repeated excitation by 700 nm (**i**) and 980 nm (**j**) LED.

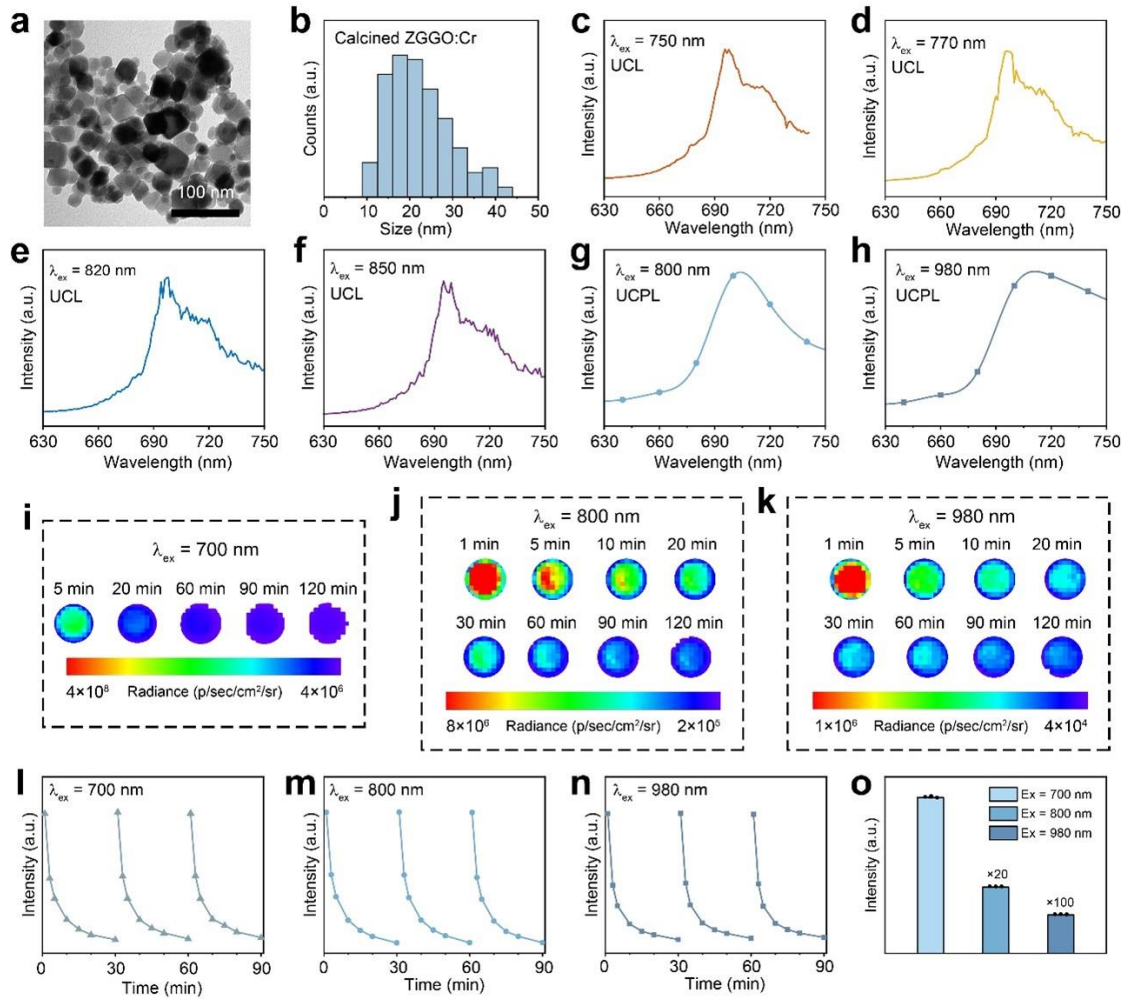

**Figure S5 UCPL in ZGGO:Cr nanoparticles.** TEM image (**a**) and size distribution (**b**) of ZGGO:Cr nanoparticles after calcination. **c-f**, UCL spectra of ZGGO:Cr nanoparticles under NIR excitation under different wavelengths. **g**, **h**, UCPL spectra of ZGGO:Cr nanoparticles under different NIR excitation. **i-k**, UCPL decay images of ZGGO:Cr nanoparticles after pre-excitation by different NIR LED. **l-n,** Reproducibility of UCPL in ZGGO:Cr after repeated pre-excitation by different NIR LED. **o**, UCPL intensity of ZGGO:Cr nanoparticles at 1 min post-excitation by 700 nm, 800 nm and 980 nm LED.

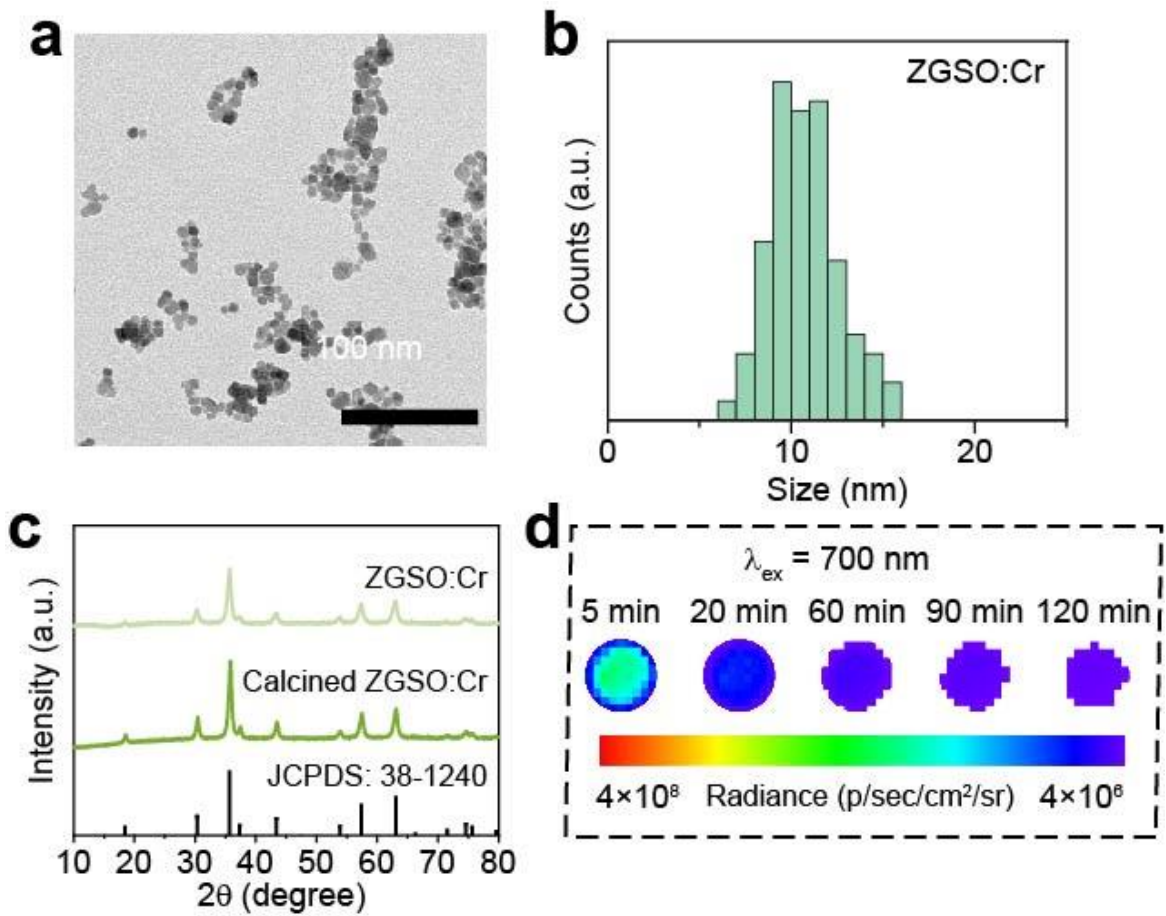

**Figure S6 Characterization of ZGSO:Cr nanoparticles.** TEM image (**a**) and size distribution (**b**) of ZGSO:Cr nanoparticles. **c**, XRD patterns of ZGSO:Cr nanoparticles before and after calcination. **d**, UCPL decay images of ZGSO:Cr nanoparticles.

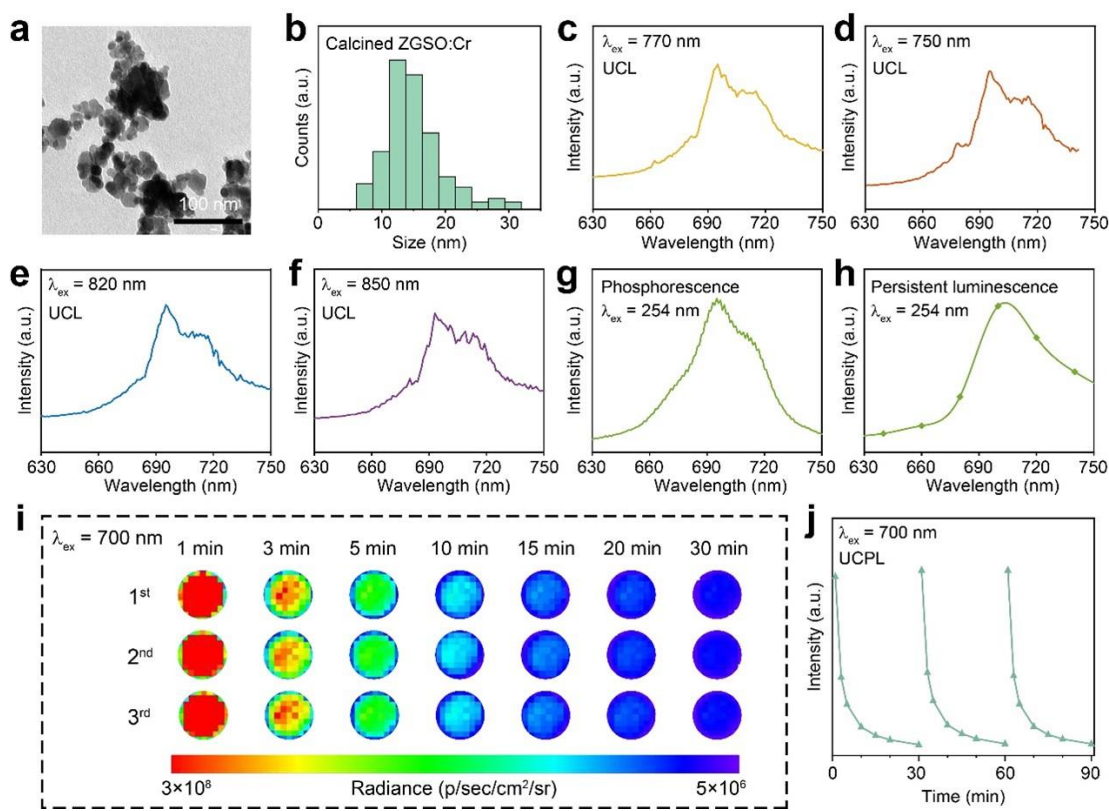

**Figure S7 UCPL in ZGSO:Cr nanoparticles.** TEM image (**a**) and size distribution (**b**) of ZGSO:Cr nanoparticles after calcination. **c-f**, UCL spectra of ZGGO:Cr nanoparticles under NIR excitation under different wavelengths. Phosphorescence (**g**) and persistent luminescence (**h**) spectrum of ZGO:Cr nanoparticles under 254 nm excitation. UCPL decay images of ZGSO:Cr nanoparticles (**i**) and the corresponding intensities (**j**) upon repeated excitation.

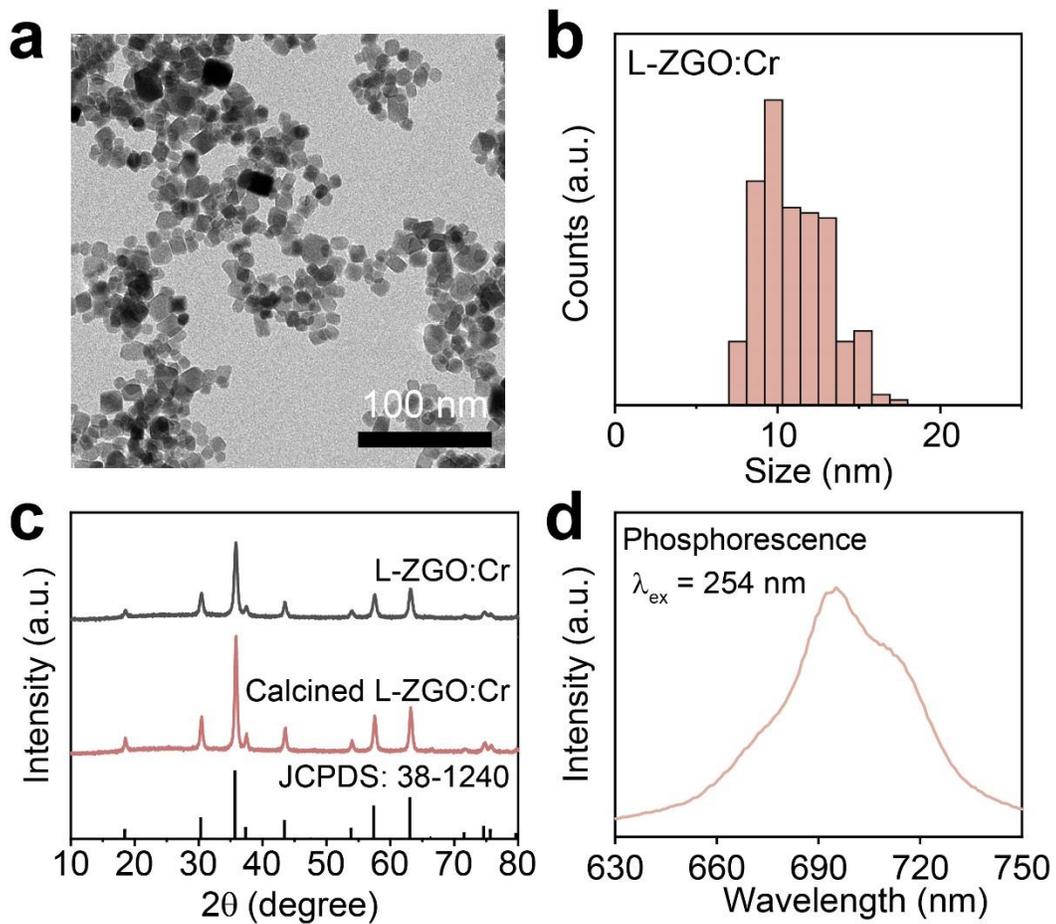

**Figure S8 Characterization of L-ZGO:Cr nanoparticles.** TEM image (**a**) and the size distribution (**b**) of L-ZGO:Cr nanoparticles. **c**, XRD patterns of L-ZGO:Cr nanoparticles before and after calcination. **d**, Phosphorescence spectrum of L-ZGO:Cr nanoparticles.

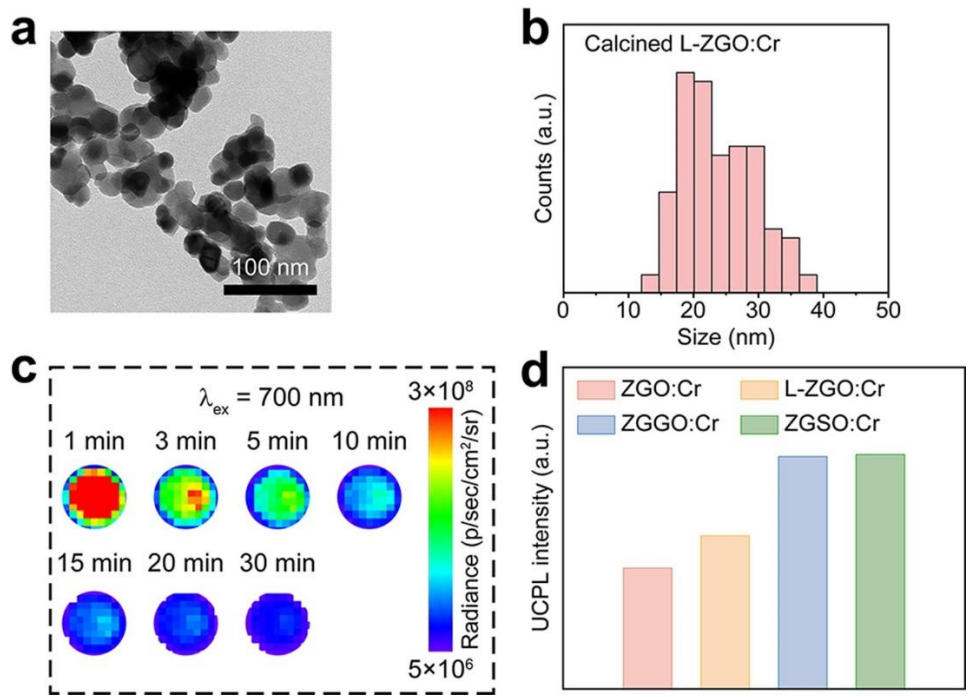

**Figure S9 UCPL in L-ZGO:Cr nanoparticles.** TEM image (**a**) and size distribution (**b**) of L-ZGO:Cr nanoparticles after calcination. **c**, UCPL decay images of L-ZGO:Cr nanoparticles after pre-excitation by 700 nm LED. **d**, UCPL intensity of ZGO:Cr, L-ZGO:Cr, ZGGO:Cr and ZGSO:Cr nanoparticles at 1min post-excitation by 700 nm LED.

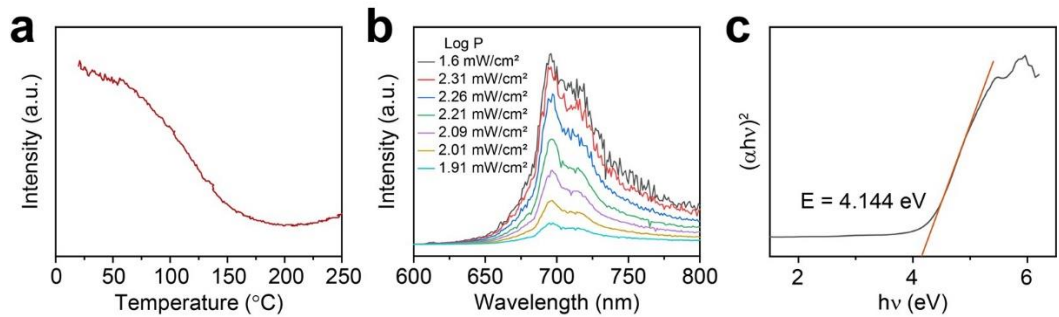

**Figure S10 Optical properties of ZGO:Cr nanoparticles. a**, Thermoluminescence spectrum of ZGO:Cr nanoparticles. **b**, UCL spectra of ZGO:Cr nanoparticles under 800 nm excitation with different power densities. **c**, Tauc plot of ZGO:Cr nanoparticles.

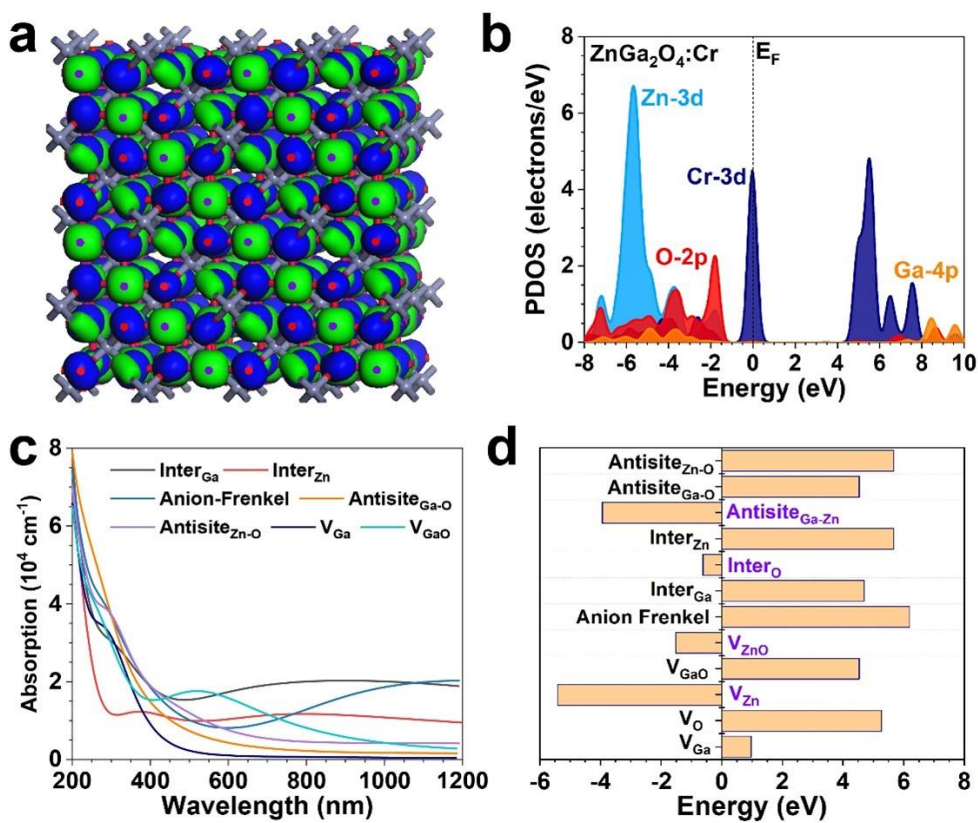

**Figure S11 DFT calculations for different properties of ZGO. a**, Thermoluminescence spectrum of ZGO:Cr nanoparticles. **b**, UCL spectra of ZGO:Cr nanoparticles under 800 nm excitation with different power densities. **c**, Tauc plot of ZGO:Cr nanoparticles.

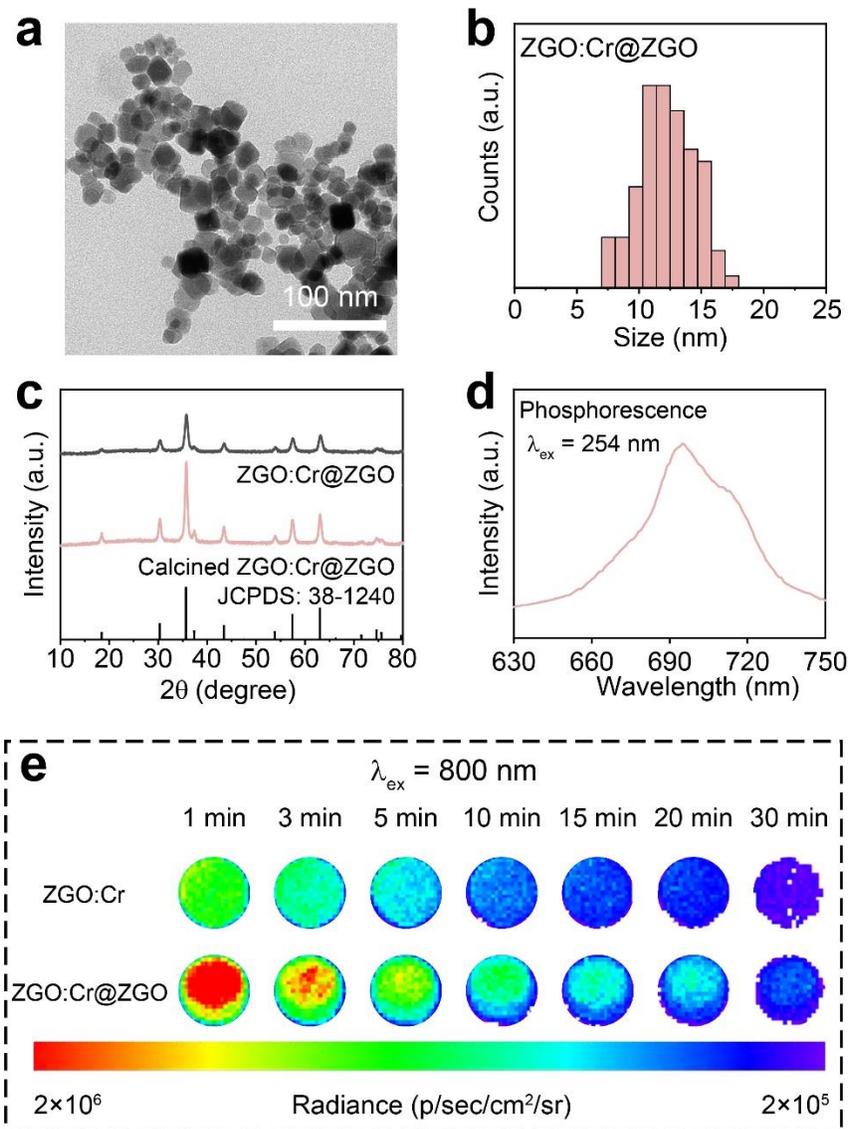

**Figure S12 Characterization of ZGO:Cr@ZGO nanoparticles.** TEM image (**a**) and the size distribution (**b**) of ZGO:Cr@ZGO nanoparticles. **c**, XRD patterns of ZGO:Cr@ZGO nanoparticles before and after calcination. **d**, Phosphorescence spectrum of ZGSO:Cr nanoparticles. **e**, UCPL decay images of ZGO:Cr and ZGO:Cr@ZGO nanoparticles.

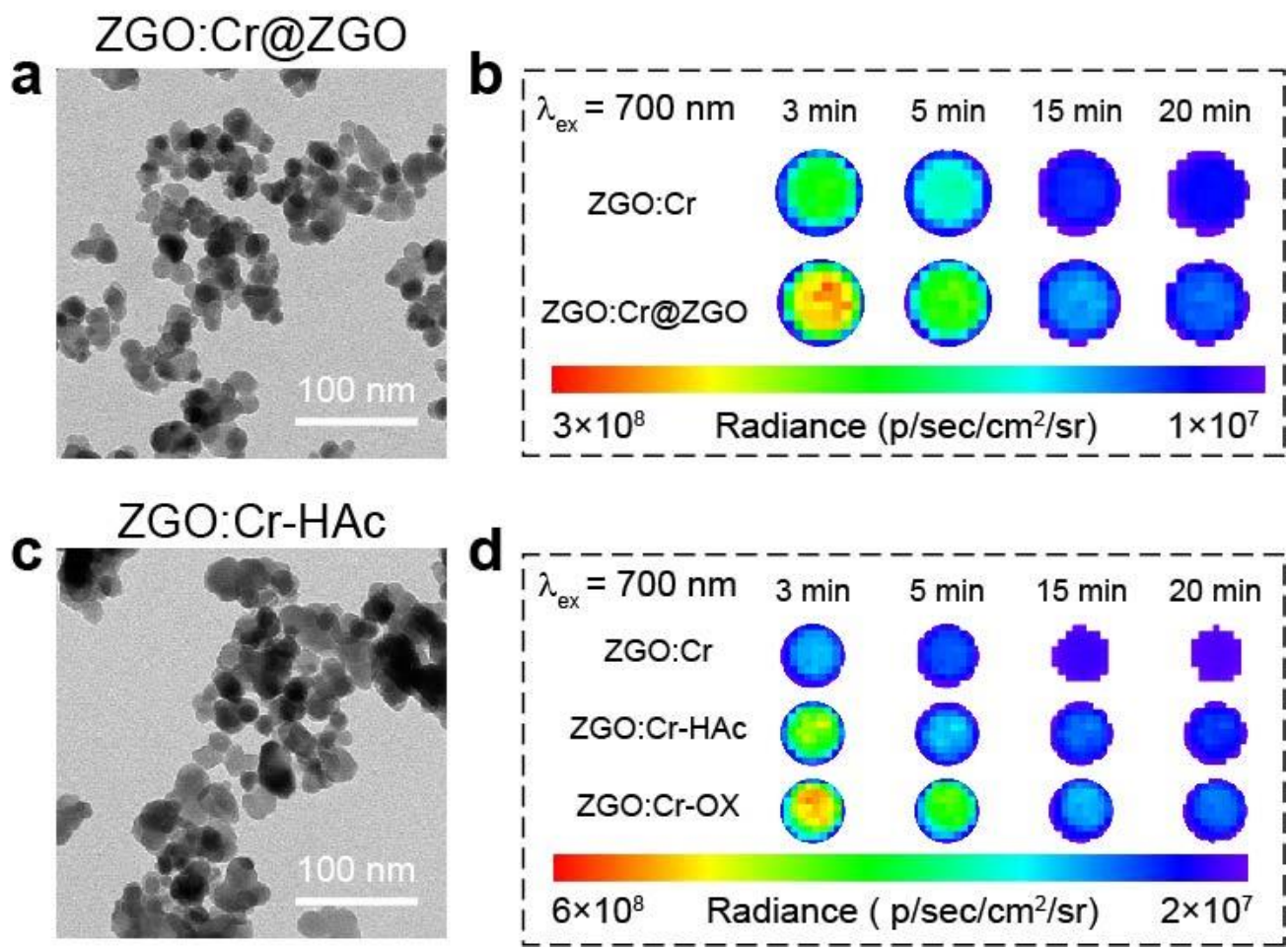

**Figure S13 UCPL in surface passivated ZGO:Cr nanoparticles.** TEM image (**a**) and UCPL decay images (**b**) of ZGO:Cr@ZGO nanoparticles. **c**, TEM image of ZGO:Cr-HAc after calcination. **d**, UCPL decay images of ZGO:Cr, ZGO:Cr-HAc, and ZGO:Cr-OX nanoparticles.

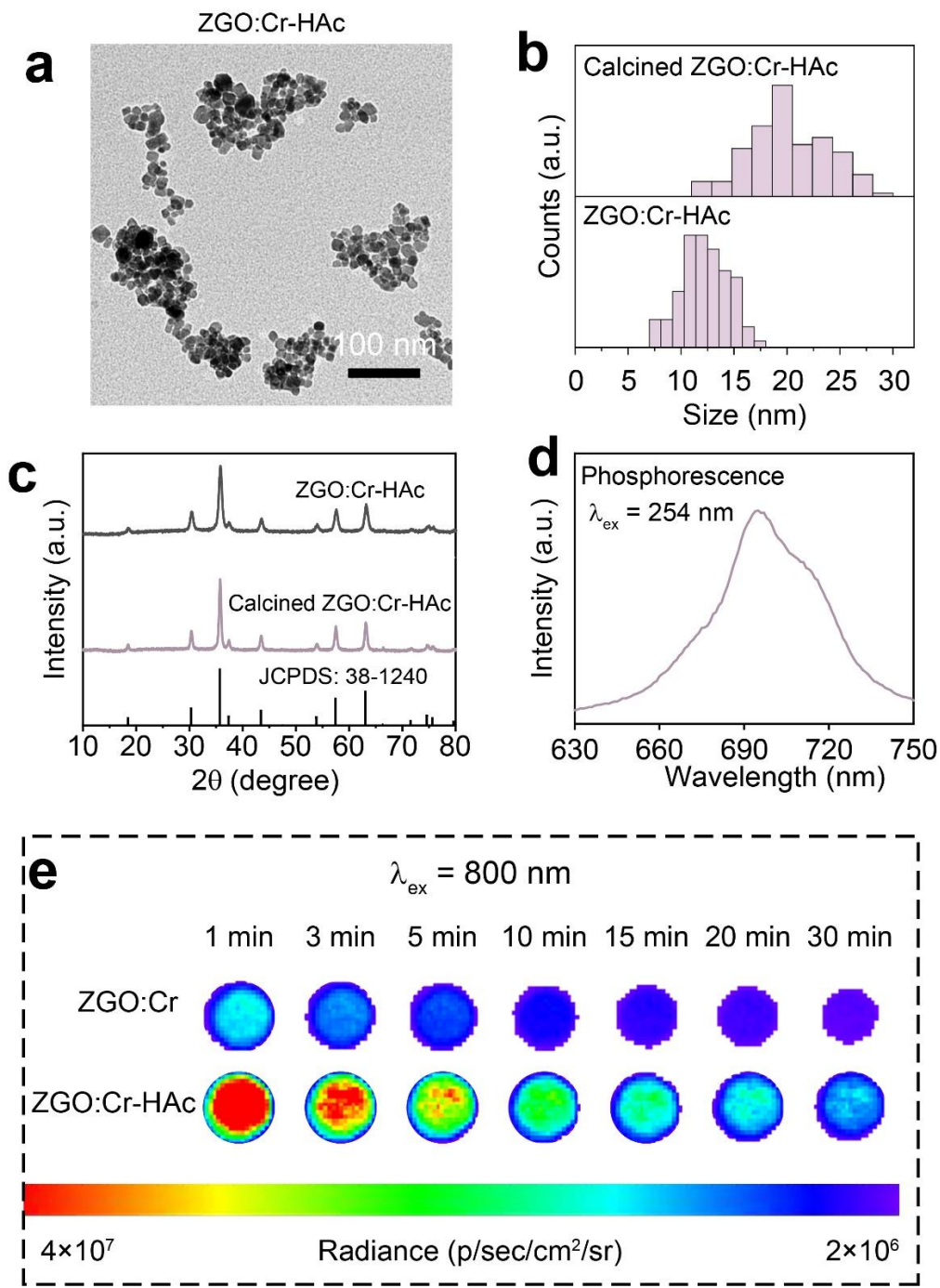

**Figure S14 Characterization of ZGO:Cr-HAc nanoparticles. a**, TEM image of ZGO:Cr-HAc nanoparticles. The size distribution (**b**) and XRD patterns (**c**) of ZGO:Cr-HAc nanoparticles before and after calcination. **d**, Phosphorescence spectrum of ZGO:Cr-HAc nanoparticles. **e**, UCPL decay images of ZGO:Cr-HAc nanoparticles.

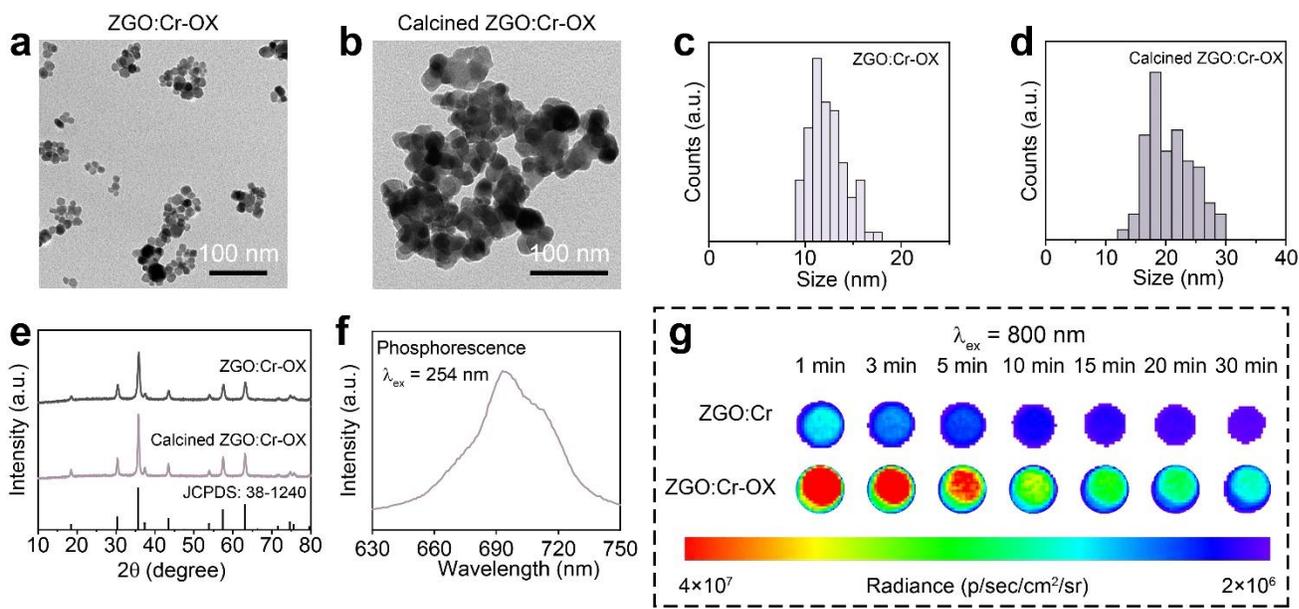

**Figure S15 Characterization of ZGO:Cr-OX nanoparticles.** TEM image of ZGO:Cr-OX nanoparticles before (**a**) and after (**b**) calcination. The size distribution of ZGO:Cr-OX nanoparticles before (**c**) and after (**d**) calcination. **e**, XRD patterns of ZGO:Cr-OX nanoparticles before and after calcination. **f**, Phosphorescence spectrum of ZGO:Cr-OX nanoparticles. **g**, UCPL decay images of ZGO:Cr-OX nanoparticles.

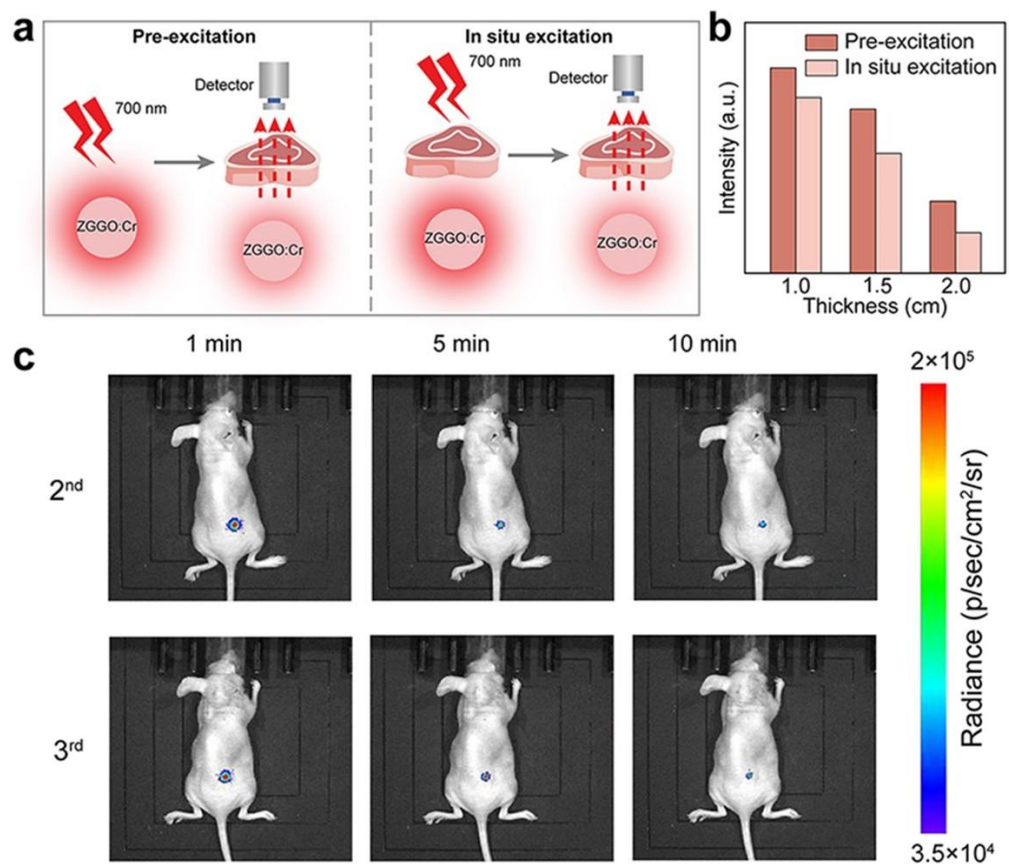

**Figure S16 UCPL imaging based on ZGGO:Cr nanoparticles. a,** Schematic illustration of UCPL imaging by pre-excitation and in-situ excitation manner. **b,** UCPL intensity of ZGGO:Cr nanoparticles at 1min post-excitation with different pork tissue coverage. **c**, UCPL images of mice bearing subcutaneously injected ZGGO:Cr upon repeated excitation by the 700 nm LED.

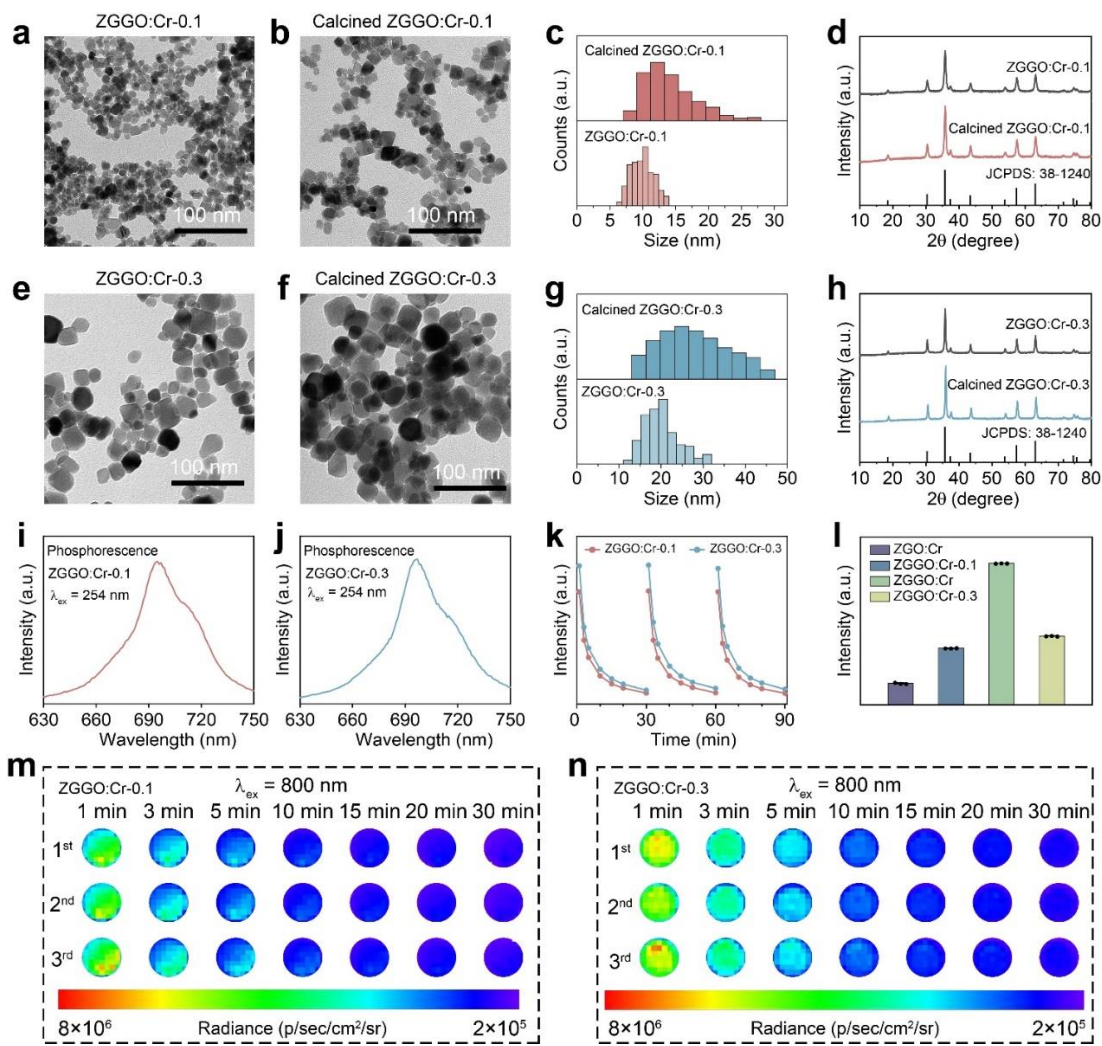

**Figure S17 Characterization of ZGO:Cr nanoparticles doped with different amounts of Ge.** TEM image of $Zn_{1.1}Ga_{1.8}Ge_{0.1}O_4$:Cr (ZGGO-0.1) nanoparticles before (**a**) and after (**b**) calcination. Size distributions (**c**) and XRD patterns (**d**) of ZGGO-0.1 nanoparticles before and after calcination. TEM image of $Zn_{1.3}Ga_{1.4}Ge_{0.3}O_4$:Cr (ZGGO-0.3) nanoparticles before (**e**) and after (**f**) calcination. Size distributions (**g**) and XRD patterns (**h**) of ZGGO-0.3 nanoparticles before and after calcination. Phosphorescence spectra of ZGGO-0.1 (**i**) and ZGGO-0.3 (**j**) nanoparticles. **k**, UCPL decay curves of ZGGO-0.1 and ZGGO-0.3 nanoparticles. **l**, Comparation of UCPL intensity at 1 min post-excitation between ZGO:Cr, ZGGO-0.1, ZGGO and ZGGO-0.3 nanoparticles. UCPL decay images of ZGGO-0.1 (**m**) and ZGGO-0.3 (**n**) nanoparticles.